# Dynamics of dissipative solitons in a high repetition rate normal-dispersion erbium-doped fiber laser

Yiyang Luo[1,2], Lei Li[1], Luming Zhao[1], Qizhen Sun[2], Zhichao Wu[1,2], Zhilin Xu[2], Songnian Fu[2], and Deming Liu[2]

[1]Jiangsu Key Laboratory of Advanced Laser Materials and Devices, School of Physics and Electronic Engineering, Jiangsu Normal University, Xuzhou 221116, Jiangsu, P. R. China;

[2]School of Optical and Electronic Information, National Engineering Laboratory for Next Generation Internet Access System, Huazhong University of Science and Technology, Wuhan 430074, Hubei, P. R. China.

This work is supported by the sub-Project of the Major Program of the National Natural Science Foundation of China (N0.61290315), the National Key Scientific Instrument and Equipment Development Project of China (No. 2013YQ16048707), and the National Natural Science Foundation of China (No. 61275004, No. 61275109). Corresponding author: L. Li (sdulilei@gmail.com).

**Abstract:** The dynamics of dissipative solitons (DSs) are explored in a high repetition rate normal-dispersion erbium-doped fiber laser for the first time. Despite of the high fundamental repetition rate of 129 MHz and thus the low pulse energy, a DS train with a dechirped pulse width of 418 fs, period-doubling of single and dual DSs, as well as 258 MHz 2nd-order harmonic mode-locking of DSs can be observed in the fiber laser with increasing pump power and appropriate settings. A transmitted semiconductor saturable absorber and a wavelength division multiplexer/isolator/tap hybrid module are employed to simplify the laser configuration, thus not only increasing the repetition rate, but also enhancing the stability and robustness of the fiber laser due to the commercial availability of all the components.

## 1. Introduction

High repetition rate mode-locked erbium-doped fiber lasers (EDFLs) have been extensively explored due to their numerous applications in optical frequency metrology, telecommunication, high-speed optical sampling and optical sensing [1-4]. Compared with actively mode-locked EDFLs [5] characterized by complicated configurations and high cost, passively mode-locked EDFLs possess distinct advantages of compactness, flexibility and improved performance, which have been demonstrated through adopting various mode-locking techniques, such as nonlinear polarization rotation (NPR) technique [6], semiconductor saturable absorber mirror (SESAM) [7], and nanomaterial (graphene, $MoS_2$, $Bi_2Te_3$, etc.) based saturable absorbers (SAs) [8-10]. Thus it can be seen that passively mode-locking technique is much more desirable to provide high repetition rate pulse trains. As a possible approach, harmonic mode-locking of fiber lasers is adopted to increase the repetition rate. In particular, a 232 MHz 10th-order passively harmonic mode-locked normal-dispersion EDFL [6] and a 2.5 GHz passively harmonic mode-locked EDFL [11] have been demonstrated, respectively. However, the inherent timing jitter and accompanying power fluctuation of the high-order harmonic mode-locking are challenging obstacles for their developments. Alternately, controlling the cavity dimensions paves another way to achieve high fundamental repetition rates in passively mode-locked EDFLs. K. Wu *et al.* has reported a 463 MHz fundamental mode-locked EDFL based on few-layer molybdenum disulfide ($MoS_2$) SA [12]. The linear cavity configuration incorporated with a $MoS_2$ SA significantly shortens the cavity length for increasing the fundamental repetition rate. Nevertheless, the few-layer $MoS_2$ SA is not commercial available and ubiquitous, thus impeding its extensively industrial applications. Moreover, this fiber laser mode-locked in the anomalous-dispersion regime generates the so-call nonlinear Schrodinger equation (NLSE) solitons characterized by a hyperbolic pulse shape and limited pulse energy [13-16].

Previous studies have demonstrated that optical solitons can be also generated in the normal-dispersion regime, the dynamics of which is governed by the complex Ginzburg-Landau equation (GLE) [17-21]. Different from the formation mechanism of NLSE solitons in the anomalous-dispersion regime, spectral filter plays significant roles in the formation of GLE solitons, namely dissipative solitons (DSs) in normal-dispersion regime. As a consequence, these DSs are characterized by the steep spectral edges and Gaussian pulse shape with strong frequency chirp, which can be easily amplified and compressed for achieving ultrashort pulses with large energy [22-26]. Thus it can be seen that high repetition rate DSs formed in the normal-dispersion regime are much more attractive than NLSE solitons formed in the anomalous-dispersion regime for their distinct advantages and promising applications. Indeed, deriving from the anomalous dispersion of single-mode fiber (SMF) at 1.55 μm, dispersion compensation fiber (DCF) or dispersion shift fiber (DSF) is usually utilized inside the cavity of the DS EDFLs to manage the net normal dispersion. Whereupon, generation of high repetition rate DSs is always excluded due to the extended cavity lengths. To address this problem, Zhang et al. has proposed an all-fiber passively mode-locked EDFL based on a tilted fiber grating (TFG) polarizer, and 250 MHz fundamental repetition rate is achieved due to the compact laser configuration [27]. Nevertheless, the mode-locker of nonlinear wave interference based NPR technique is intrinsically environmental unstable and high pump power is needed to initialize the mode-locking, which are not appropriate for practical applications.

In this paper, we experimentally develop a high repetition rate passively mode-locked EDFL operated in the normal-dispersion regime. All the components are commercially available. A transmitted semiconductor saturable absorber (SESA) and a wavelength division multiplexer/isolator/tap hybrid module (WDM/Isolator/Tap hybrid module) play crucial roles in shortening the cavity length for achieving a high fundamental repetition rate. Benefiting from the large normal dispersion of the EDF and the compact laser configuration, the proposed passively mode-locked EDFL can be managed to operate in the normal-dispersion regime without adopting any other dispersion-compensated components. Multiple-state DSs are experimentally demonstrated for the first time in this high repetition rate normal-dispersion EDFL. Particularly, fundamental mode-locking, period-doubling, and 2nd-order harmonic mode-locking are respectively observed with increasing pump power and appropriate settings. Moreover, the laser design of high repetition rate EDFLs is also discussed, revealing that the proposed EDFL is characterized by ubiquity and the repetition rate could be potentially scaled up.

## 2. Experimental Setup

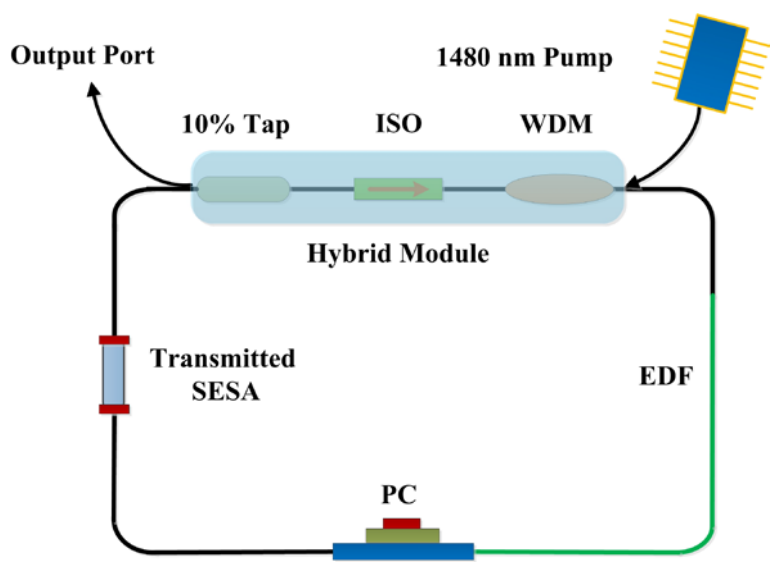

Fig. 1. Schematic of the passively mode-locked EDFL.

The schematic of the laser setup is depicted in Fig. 1. The mode-locked fiber laser possesses a compact ring-cavity configuration. A 0.75 m erbium-doped fiber (EDF, OFS EDF80) with group-velocity dispersion (GVD) parameter of −32 (ps/nm)/km is used as the gain medium, and pumped by a high power 1480 nm Raman fiber laser source. A WDM/Isolator/Tap hybrid module (1480/1550 nm WDM, polarization independent isolator, and 10% output tap) is incorporated to shorten the cavity length. A commercial available transmitted SESA (BATOP, SA-1550-25-2ps-FC/PC) is adopted to initialize the mode locking of the fiber laser. The SA (GaAs) chip with a thickness of 100 μm is fiber butt-coupled inside a ceramic ferrule, and connected to single-mode fiber (SMF) patch cables with FC/PC connectors. The SESA possesses large absorption of 25%, modulation depth of 15%, non-saturable loss of 10%, saturation fluence of 300 μJ/cm$^2$, damage threshold of 100 MW/cm$^2$, and recovery time of 2 ps. An in-line fiber polarization controller (PC) is applied between the EDF and the transmitted SESA to slightly adjust the intra-cavity polarization. All the pigtails of the components are SMF with GVD parameter of 16 (ps/nm)/km, particularly including 0.34 m for the WDM/Isolator/Tap hybrid module and 0.51 m for the transmitted SESA. Consequently, the overall length of the fiber laser is about 1.60 m with a net normal dispersion of about 0.013 ps$^2$, thus leading to the formation of high repetition rate DSs. In addition, an optical spectrum analyzer (OSA, Yokogawa AQ6370C-20) with a resolution of 0.02 nm and a 1 GHz real-time oscilloscope (OSC, Agilent DSO9104A) together with a 2 GHz photodetector (PD, Thorlab DET01CFC) are respectively employed to monitor the optical spectra and the pulse train of DSs. The radio frequency (RF) spectrum is analyzed by a 3 GHz electrical spectrum analyzer (ESA, Agilent N9320B), and the pulse width is measured by a commercial autocorrelator (Femtochrome FR-103HS).

## 3. Experiment and Discussions

Due to the saturable absorption of the transmitted SESA, self-started mode-locking of the EDFL can be easily achieved, and the pulses are shaped into DSs with a rectangular spectral shape due to the net normal dispersion. Through tuning the pump power to 158 mw and appropriately adjusting the intra-cavity polarization, fundamental mode-locking of DSs is firstly obtained as illustrated in Fig. 2. The optical spectrum of the DSs is shown in Fig. 2(a), exhibiting steep spectral edges with a 3-dB bandwidth of 10.2 nm. The slightly modulated spectrum top is most likely caused by the effective gain which is the combined effect from gain, gain dispersion, component filter etc. The autocorrelation trace depicted in Fig. 2(b) declares a pulse width of 5.0 ps if a Gaussian pulse shape is assumed. Thus, the time-bandwidth product of the DSs is 6.4, indicating that these solitons are strongly chirped. Figures. 2(c) and (d) respectively show the oscilloscope trace and radio frequency (RF) spectrum of the DSs. The fundamental repetition rate is fixed at 129 MHz and the pulse interval is 7.7 ns which agree well with the cavity length. The average output power of the fundamental mode-locking is around 5.9 mw with the pulse energy of 45.5 pJ. The overall laser efficiency is around 3.7%. It should be noted that the low laser efficiency is principally attributed to the relatively large insertion loss induced by the SESA with absorption of 25%.

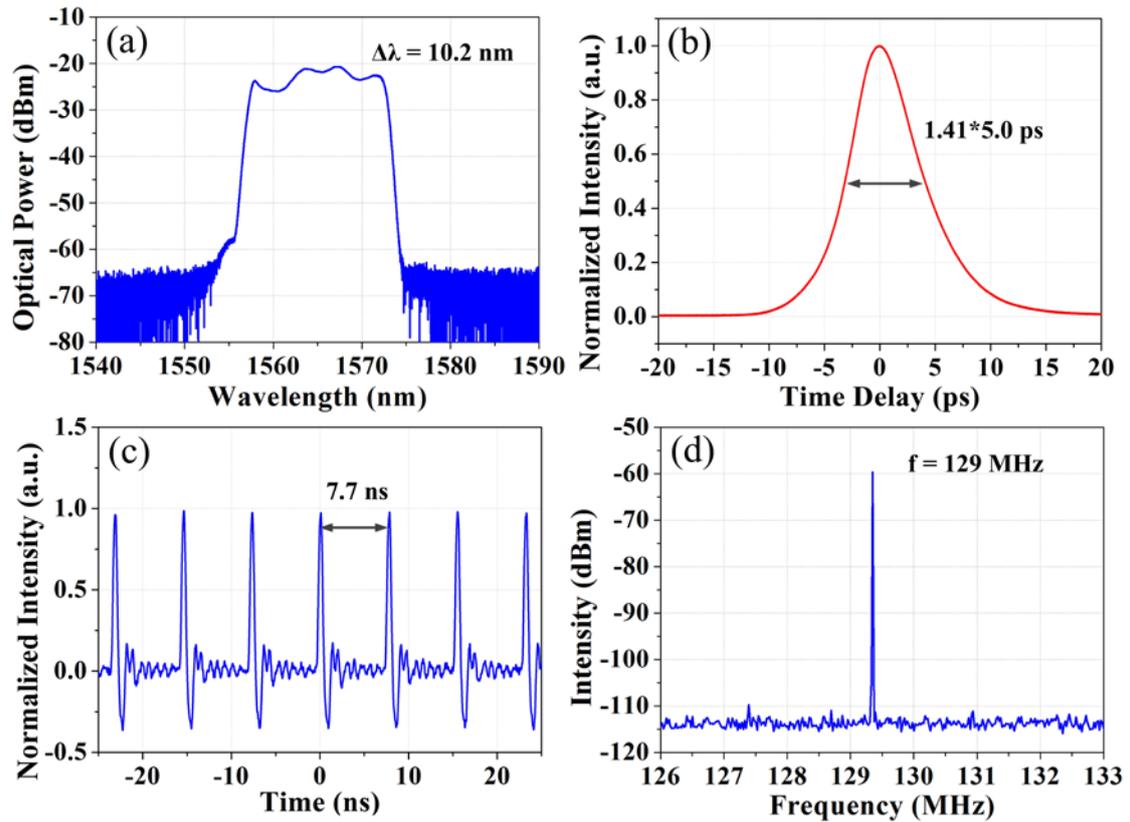

Fig. 2. (a) Optical spectrum, (b) autocorrelation trace, (c) oscilloscope trace, and (d) RF spectrum of the 129 MHz fundamental mode-locked DSs.

It is well known and also experimentally revealed that DSs formed in the normal-dispersion regime are strongly chirped. Therefore, SMF out-side the laser cavity can be utilized to dechirp the output solitons for obtaining an ultrashort pulse. Figure. 3 shows the autocorrelation traces of the chirped and dechirped pulses (dashed blue line and solid red line, respectively). Particularly, the pulse width can be compressed to 418 fs with the assumption of a Gaussian pulse shape following the cut-back method. And no obvious difference is observed for the optical spectra between the original DSs and the dechirped ones. After pulse compression, the minimum time-bandwidth product is about 0.53, slightly larger than the transform-limited value of 0.44. The experimental result of the pulse compression demonstrates once again that the high repetition rate DSs are strongly chirped. Moreover, it should be noted that, the autocorrelation trace of the dechirped pulse exhibits small pedestals, which derives from the nonlinear chirp accumulated when the pulse propagates in the laser cavity.

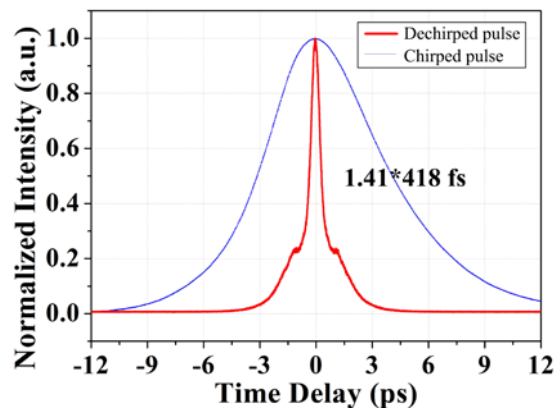

Fig. 3. Autocorrelation traces of the chirped and dechirped pulses.

Next, starting from the stable fundamental mode-locked operation state, we gradually increase the pump power and carefully adjust the PC, thus leading to an increasing peak power of the pulses. To a certain point with the pump power of 170 mw, the pulse intensity of the DSs as shown in Fig. 4 becomes no longer identical, but alternates between two different values, and repeats every two round-trips, namely forming the so-called period-doubling. Generally, high repetition rate pulses are characterized by a relatively low pulse peak power, which is not beneficial for the generation of period-doubling. Whereas in the dispersion-managed fiber laser with net normal dispersion, the chirped pulses can be compressed in the anomalous-dispersion SMF pigtails after amplified in the EDF. Therefore, the stronger the pump power, the larger is the chirp accumulated in the EDF, and consequently narrower pulse with higher peak power. Once the pulse peak power is beyond a certain value, the period-doubling of DSs occurs [28]. Our experimental results reveal that despite of the largely chirped pulses and a high repetition rate greater than 100 MHz, the period-doubling of DSs can be still observed in the normal-dispersion regime.

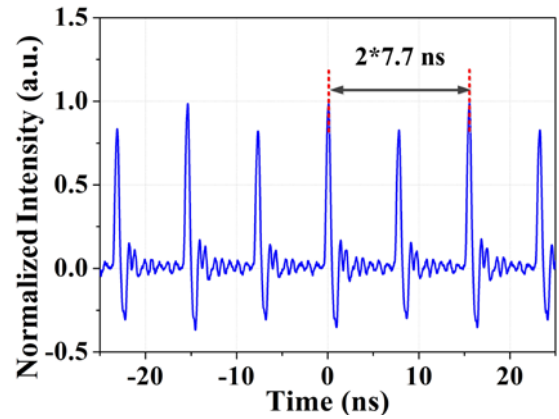

Fig. 4. Oscilloscope trace of the period-doubling of DSs.

Furthermore, we continue increasing the pump power from the period-doubling state of DSs. Owing to the peak power clamping effect [29], the fundamental mode-locked DS pulses cannot be sequentially amplified, instead tending to operate at multiple-soliton state. As expected, the multiple-soliton state is obtained accompanying with a decreased pulse peak power. When the pump power is set at 181 mw with proper PC setting, these DSs rearrange themselves in a regular position, namely achieving the 2nd-order harmonic mode-locking of the EDFL in the normal-dispersion regime. The optical spectrum of the 2nd-order harmonic mode-locked DSs is depicted in Fig. 5(a) with a 3-dB bandwidth of 7.5 nm. Figure. 5(b) illustrates the autocorrelation trace of the DSs, and the pulse width is estimated to be 4.9 ps with the assumption of a Gaussian pulse shape. The time-bandwidth product is calculated as 4.6, indicating that these DSs are strongly chirped as well. Figure. 5(c) shows the oscilloscope trace of the DSs, where 2 pulses coexist in one cavity roundtrip time of 7.7 ns with an equal interval, corresponding to a repetition rate of 258 MHz as shown in the inset of Fig. 5(c). Meanwhile, the average output power is increased to 6.6 mw, while the individual pulse energy is inevitably reduced to 25.5 pJ due to the reduplicated repetition rate. Through slightly adjusting the intra-cavity polarization state, dual-soliton state are observed as well. The pulse intensity of each of the two co-existed DSs is no longer identical, but repeats every two round-trips, evidently manifesting the generation of period-doubling as shown in Fig. 5(d). In the current state, the pulse energy is almost half of that of Fig. 4. However, the phenomena of period doubling can be still obtained. The detailed relationship between the pulse energy and the appearance of the period doubling will be discussed elsewhere. To our best knowledge, period-doubling of high repetition rate dual DSs is observed for the first time.

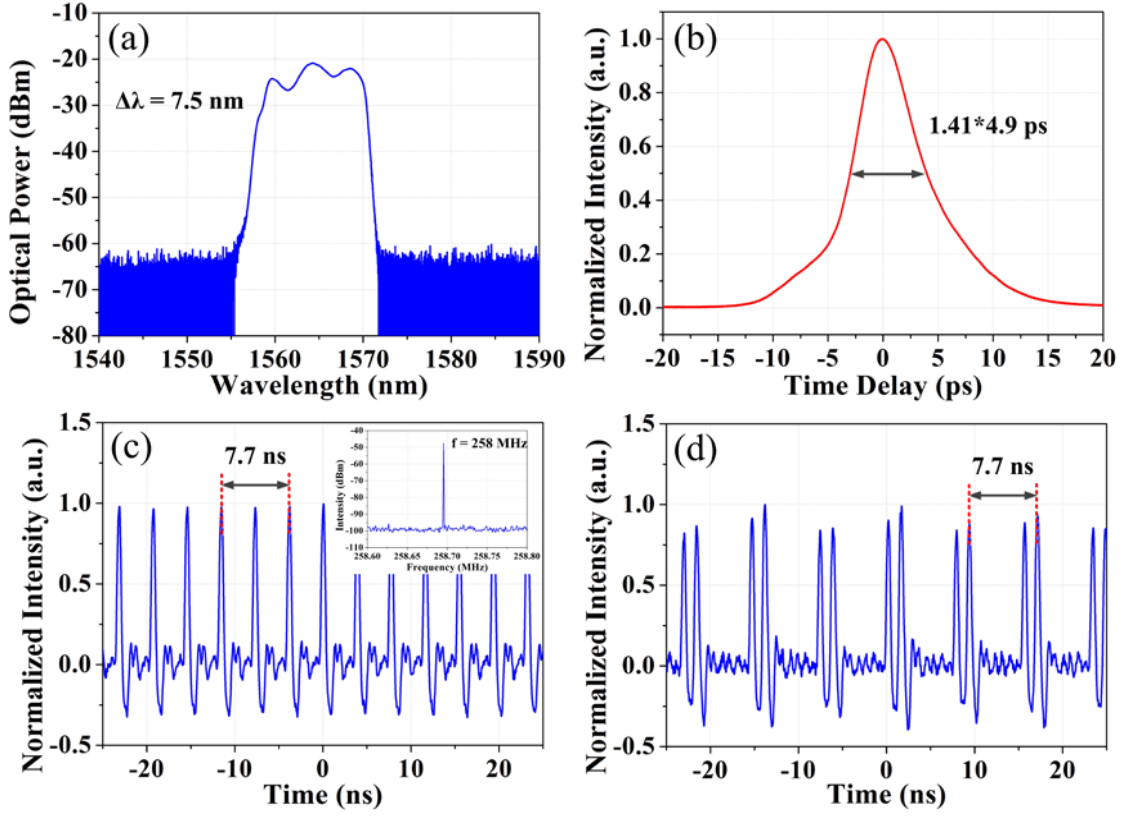

Fig. 5. (a) Optical spectrum, (b) autocorrelation trace and (c) oscilloscope trace of the 2nd-order harmonic mode-locked DSs (inset of Fig. 5(c) is the corresponding RF spectrum); (d) oscilloscope trace of the period-doubling of dual DSs.

For practical applications, there are two main challenges existing in the implementation of a stable and robust EDFL with a high repetition rate. Firstly, high fundamental repetition rate is straightforwardly ascribed to a compact configuration of the EDFLs. And secondly, a suitable mode-locking mechanism with moderate pump power requirements is crucial as well [30]. From our perspective, the proposed passively mode-locked EDFL could well address the aforementioned two problems. On one hand, the laser configuration is simplified by the transmitted SESA and the WDM/Isolator/Tap hybrid module for achieving a high fundamental repetition rate, as well as spliced together for enhancing the robustness of the EDFL. Moreover, all the utilized components are commercial available with a mature technology and long-term reliability, thus making the EDFL ubiquitous and widely applicable. On the other hand, compared with the NPR technique of intrinsically unstable, the transmitted SESA is intrinsically stable and experimentally repeatable. Additionally, the pump power requirement (below 200 mW) of this reported EDFL is much lower than that in NPR based mode-locked fiber lasers. Thus it can be seen that the transmitted SESA are much more desirable than the NPR technique to provide stable high repetition rate DSs for practical applications. Promisingly, with modest adjustment, the achievable high fundamental repetition rate of the EDFL could be increased to more than 400 MHz, which is about an overall cavity length of 50 cm. For example, we could further reduce the pigtails of the components, and use shorter EDF with higher doping. Besides, an all-polarization maintaining configuration could be also designed instead of the non-polarization maintaining one for achieving a stable linear output polarization of the DSs and an environment-resistant fiber laser.

## 4. Conclusion

We report on the first experimental observation of multiple-state DSs in a high repetition rate normal-dispersion EDFL. The laser configuration is properly simplified for achieving a high fundamental repetition rate through using a transmitted SESA and a WDM/Isolator/Tap hybrid module. In particular, 129 MHz fundamental mode-locking of DSs with a dechirped pulse width of 418 fs, period-doubling of single and dual DSs, as well as 258 MHz 2nd-order harmonic mode-locking of DSs are respectively observed in this EDFL. The observations greatly enrich our understanding towards the dynamics of DSs in a high repetition rate fiber laser. The repetition rate of the EDFL could be potentially scaled up by further shortening the cavity length, and the stability could be optimized by adopting an all-polarization designed laser cavity.